\documentclass[RNAAS]{aastex62}

\begin{document}

\title{A Method for a Pseudo-Local Measurement of the Galactic Magnetic Field}

\correspondingauthor{Steven Spangler}
\email{steven-spangler@uiowa.edu}

\author[0000-0002-4909-9684]{Steven R. Spangler}
\affiliation{Department of Physics and Astronomy, University of Iowa}

\begin{abstract}
Much of the information about the magnetic field in the Milky Way and other galaxies comes from measurements which are path integrals, such as Faraday rotation and the polarization of synchrotron radiation of cosmic ray electrons.   The measurement made at the radio telescope results from the contributions of volume elements along a long line of sight.  The inferred magnetic field is therefore some sort of average over a long line segment.  A magnetic field measurement at a given spatial location is of much more physical significance.  In this paper, we point out that HII regions fortuitously offer such a ``point'' measurement, albeit of one component of the magnetic field, and averaged over the sightline through the HII region.  However, the line of sight (LOS) through an HII region is much smaller (e.g. 30 - 50 pc) than one through the entire Galactic disk, and thus constitutes a ``pseudo-local'' measurement. We use published HII region Faraday rotation measurements to provide a new constraint on the magnitude of magnetohydrodynamic (MHD) turbulence in the Galaxy, as well as to raise intriguing speculations about the modification of the Galactic field during the star formation process.  
\end{abstract}

\section{1. Introduction} 
It has been known for decades that the interstellar medium is threaded by a magnetic field.  This is not surprising, since almost all phases of the interstellar medium (ISM) have sufficient level of ionization to satisfy the conditions for a plasma parameter $\Lambda_e \gg 1$ \citep[number of particles per DeBye sphere large,][]{Nicholson83}, and thus constitute plasmas.  Currents flowing in these plasmas generate a magnetic field. Measurements that provide information on the magnitude and functional form of the Galactic field include Faraday rotation of radio sources, both in the Galaxy and beyond, Zeeman splitting of magnetically-sensitive transitions of the hydrogen atom and the hydroxyl, methanol and water molecules, polarization of the Galactic nonthermal synchrotron radiation, and polarization of visible and infrared light due to the alignment of interstellar grains in the Galactic magnetic field.  Recent reviews on the Galactic field include \citet{Ferriere11} and \citet{Han17}.  

Studies extending back over several decades indicate that the Galactic field consists, in part, of a Galaxy-wide, large scale field that can be described, at least empirically, by analytic functions of galactocentric coordinates, and a superposed, spatially-random component that is plausibly interpreted as magnetohydrodynamic (MHD) turbulence, presumably similar to that which exists in the solar wind. An example of a study that prescribes the form of the large scale, ``deterministic'' component of the field is \citet{vanEck11}.  Reviews which consider the state of knowledge of both the large scale and turbulent components are given in \citet{Ferriere11} and \citet{Han17}.  A very recent review which summarizes what we know and can speculate about the turbulent component is \citet{Ferriere20}.   

It is probably safe to say that all observational investigations have concluded that the turbulent component is comparable in magnitude to the large scale component.  The exact value of the magnitude of the turbulent component depends on what is assumed for the outer scale of the turbulence.  Values for the ratio $\frac{\delta b}{B_0}$ where $\delta b$ is the rms value of the turbulent component and $B_0$ is the magnitude of the large scale component, range from a few tens of percent to a number in excess of unity.  For example, \citet{Ferriere11} and \citet{Ferriere20} review our knowledge of the Galactic field, and pay attention to the relative magnitudes of the regular and turbulent fields, citing results of $\frac{\delta b}{B_0} \geq 3$ from two independent investigations.  Interestingly, \citet{Beck16}, in a review of what is known about magnetic fields in external galaxies similar to the Milky Way, argues that the turbulent component is substantially larger than the large scale, deterministic component.  In Section 6 of his review, Beck examines the state of knowledge of the Galactic field in the context of results for external galaxies.  He concludes that the amplitude of isotropic magnetic fluctuations is $5 \mu G$, that of anisotropic fluctuations $2 \mu G$, and that of the regular component (ordered over Galactic scales) is $2 \mu G$.  It should be emphasized that these estimates are dependent on assumptions on the outer scale of the turbulence, i.e. the scale which distinguishes a Galactic-scale ``deterministic'' component from a turbulent component.  Nonetheless, these reviews make clear that there is substantial observational evidence for a turbulent component to the Galactic magnetic field that is comparable in magnitude, or even larger than an ordered, large scale field. In this paper, we will consider the dimensionless turbulent amplitude $\frac{\delta b}{B_0}$ to be a free parameter, subject to the observational constraints cited above. Knowledge of the magnitude and spectrum of the Galactic MHD turbulence is important to Galactic astrophysics, and astrophysics in general, since this turbulence contributes to processes such as heating of the ISM through dissipation of turbulence, confinement and propagation of the cosmic rays, turbulent transport of heavy elements, and modification or regulation of the star formation process.  

One of the difficulties in determining the relative roles of a large scale and a turbulent component of the Galactic field is the fact that some of the main diagnostic techniques such as  Faraday rotation, polarization of Galactic synchrotron radiation, and dust polarization consist of path integrals over long lines of sight through the Galaxy.  What this means is that we get some sort of average of the Galactic field over a long line of sight (LOS) through the Galaxy, rather than the value at a set of points along that line of sight.   Zeeman splitting of OH and methanol maser transitions are not subject to this limitation, since the region of emission is very localized \citep{Crutcher19}.  However, as discussed in \citet{Crutcher19} regions of maser emission have very high densities compared to the general ISM, and are products of the star formation process.  We will return to this point in Section 4.  

In this paper, we discuss a type of observation which utilizes the Faraday rotation technique to obtain an estimate of the Galactic field (strictly speaking one component of the field) in a localized region, rather than an entire line of sight through the Galaxy.  The localized region is comprised of the volume of an HII region.  In this sense, the technique provides a ``local'' measurement.  In what follows, we first summarize the basis of the technique of Faraday rotation, then discuss recent observational results which allow it to be used in a ``local'' measurement of the Galactic field.  

Faraday rotation is a fundamental plasma process which consists of rotation in the plane of polarization of radio waves due to the presence of a magnetized plasma \citep{Nicholson83}. It is used as a diagnostic technique in laboratory plasmas as well as in radio astronomy. In the present context, it is applied to radio waves emitted by extragalactic radio sources and which propagate through the plasma of the interstellar medium (ISM).  The change in the position angle relative to what would be measured in the absence of the plasma, $\Delta \chi$, is given by
\begin{equation}
\Delta \chi = \left[  \left( \frac{e^3}{2 \pi m_e^2 c^4} \right) \int_L n \vec{B} \cdot \vec{ds} \right] \lambda^2
\end{equation}
In Equation (1) $e, m_e, \mbox{ and } c $ are the customary fundamental physical constants, $n$ is the electron density, $\vec{B}$ is the vector magnetic field, both functions of position along the line of sight, and $\vec{ds}$ is an increment in the line of sight from the source to the radio telescope.  Faraday rotation is a path integral measurement rather than a point measurement, so there are contributions to plasma throughout the ISM along the line of sight.  The quantity in square brackets in Equation (1) is termed the rotation measure (RM), and is usually quoted in SI units of radians/m$^2$. For most lines of sight through the ISM, the magnetic field measured is not only an average over a long line through the ISM, it is also an average weighted by the plasma density.  Information on many lines of sight around the sky is used to obtain estimates of the large scale structure of the Galactic field, as well as the contribution of the turbulent component.  

The observational impetus for this paper is a set of studies of the HII regions Rosette Nebula and W4 by A. Costa and coworkers \citep{Savage13,Costa16,Costa18}.  Costa and co-workers carried out Faraday rotation measurements with the Very Large Array (VLA) of background radio sources viewed through these two HII regions.  The main results of these investigations were: (1) rotation measures (RM) for sources viewed through the HII regions were typically much larger than for nearby sources whose lines of sight probed the Galaxy, but not the HII regions, (2) although there were large variations in the RM from source-to-source when the LOS passed through the HII region, there was a well-defined average RM (magnitude and sign) for the HII region, particularly the Rosette, (3) the sign of the HII region-induced RM (which dominated the total RM) was consistent with the sign of the large scale Galactic field in that part of the sky.  The RM due to the Rosette Nebula (Galactic longitude = 206.5 degrees) was dominantly positive, while that for W4 (Galactic longitude = 135 degrees) was dominantly negative.  

 The important point is that the magnetic field in the ``RM anomaly'' associated with the HII region is confined to the volume of the HII region rather than being an average over a line of sight of many kiloparsecs in the Galactic plane.  The radii of both HII regions are approximately 20-25 parsecs, so the magnetic field is ``local'' in this sense.   As discussed in the text earlier in this section, in view of the apparent large amplitude, turbulent component of the Galactic field, a reasonable a-priori expectation would be that the magnetic field at a point in the Galactic plane, and averaged over a volume of radius 20 - 25 parsecs, would have poor correlation with the ordered, large scale Galactic field.  The observations of the Rosette Nebula and W4 do not bear out this expectation.  In this paper, we explore the consequences of this result for our inference about the Galactic magnetic field.

\section{2. Additional Observational Results on Faraday Rotation through HII Regions} 

Two measurements is, of course, very poor statistics, even by the base standards of astronomy.  There is a need for additional measurements of the sort presented by Costa and collaborators.  Such measurements were provided by the observations of \citet{HarveySmith11}, who carried out similar observations, for similar goals, of somewhat larger HII regions like the $\lambda$ Orionis HII region. In collecting HII regions from the sample of \citet{HarveySmith11}, we restricted attention to objects in the outer two quadrants of the Galactic plane, i.e. those with $90^{\circ} \leq l \leq 270^{\circ}$.  \citet{HarveySmith11} reported the the average value of the line-of-sight component of the magnetic field $<B_{\parallel}>$ for each HII region, and this quantity was also available from the analyses of \citet{Savage13,Costa16,Costa18}.  

In the studies of Costa and coworkers, the mean LOS component of the field, $<B_{\parallel}>$, was not explicitly measured; analyses of the data concentrated on fitting plasma shell models to the sets of RM values for each HII region.  In addition to other parameters, these models were dependent on the magnitude and orientation of the magnetic field outside the HII region (see Section 5.1 of \citet{Costa16}).  A limiting type of model was one in which there was no amplification of the external magnetic field in the HII region (referred to by Costa as the ``Harvey-Smith model'').  In this case, the model fits give exactly the same quantity as the parameter $<B_{\parallel}>$ in  \citet{HarveySmith11}.  We took values from Table 4 of \citet{Costa16} and Table 6 of  \citet{Costa18} to obtain values of $<B_{\parallel}>$, which should be directly comparable to this quantity in  \citet{HarveySmith11}.  Once again,  $<B_{\parallel}>$ has a sign and a magnitude.  The values of  $<B_{\parallel}>$ from \citet{HarveySmith11}, \citet{Costa16}, and \citet{Costa18} are plotted in Figure 1, and given in Table 1. 

\begin{figure}[h!]
\begin{center}
\includegraphics[scale=0.85,angle=0]{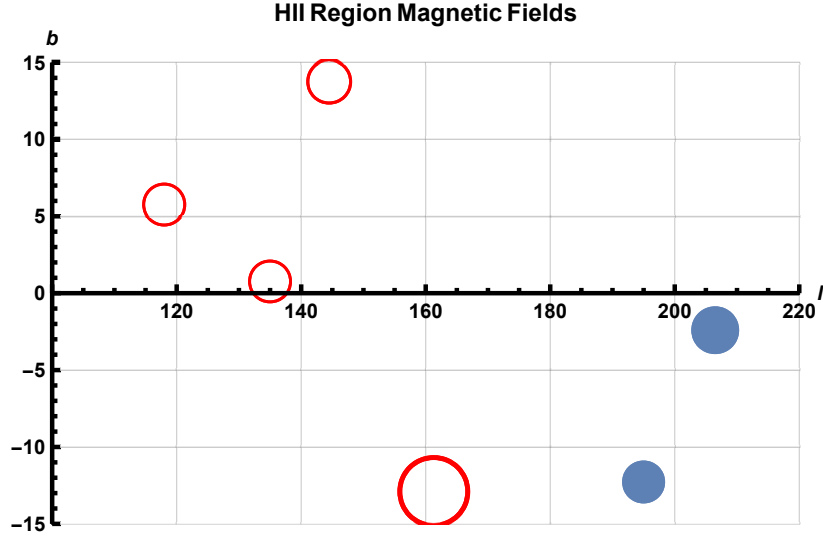}
\caption{Measurements of $<B_{\parallel}>$ from Faraday rotation measurements through 6 HII regions in the 2nd and 3rd Galactic quadrants.  Solid (blue) symbols indicate $<B_{\parallel}> >0$ (Field pointing towards observer), and open (red) symbols indicate $<B_{\parallel}> < 0$ (field pointing away from observer). The size of the plotted symbol is proportional to $\sqrt{|<B_{\parallel}>|}$. The locations of the 6 sources used are plotted in Galactic coordinates.  In the simplest case of a purely azimuthal field, the projection of the large scale Galactic field on the LOS would change sign at $l = 180^{\circ}$. In all cases, the polarity of  $<B_{\parallel}> $ is the same as that for the large scale Galactic field in that quadrant.}
\end{center}
\end{figure}

\begin{deluxetable}{|c|c|c|c|c|}
\caption{HII Region Average Magnetic Field Components}
\tablehead{\colhead{HII region}&\colhead{l}&\colhead{b}&\colhead{$<B_{\parallel}> \mu G$}&\colhead{Ref.}}
\startdata
Rosette & 206.5 & -2.1 & +2.73& \citet{Costa16} \\
W4 & 135 & +1.0 & -2.29& \citet{Costa18} \\
Sh2-264 & 195 & -12 & +2.2 & \citet{HarveySmith11} \\
Sivan 3 & 144.5 & +14 & -2.5 &  \citet{HarveySmith11} \\
Sh2-171 & 118.0 & +6 & -2.3 &  \citet{HarveySmith11} \\
Sh2-220 & 161.3 & -12.5 & -6.3 &  \citet{HarveySmith11} \\
\enddata
\end{deluxetable}
The clear result from Figure 1 is that for all 6 HII regions, the ``local'' magnetic field defined by the average over the volume of the HII region, has the same polarity as the Galactic-scale, mean magnetic field.  In none of these 6 cases are fluctuations on scales of a few tens of parsecs capable of making the ``local'' polarity opposite to the average over several kiloparsecs. 
 
\section{3. Statistics of a Local, Line of Sight Measurement of $\vec{B}$ in a Turbulent Medium}
In this section, we discuss the statistics of the LOS component of the magnetic field at a given point, when the magnetic field consists of the vector sum of a large scale, ordered magnetic field $\vec{B_0}$, and a superposed, spatially random turbulent field $\vec{\delta b}$.  For ease of following the discussion, the Appendix gives a glossary of mathematical symbols and variables used in the following.  
\subsection{Geometry of the Line of Sight and the Galactic Magnetic Field}
The geometry of the situation is shown in Figure 2.  The z axis is defined by the path from the radio source through the HII region and to the observer, with the positive direction towards the observer.  The x axis is in the plane of the sky, and points toward decreasing galactic longitude.  We also simplify the situation by adopting the good approximation that the x-z plane coincides with the Galactic plane.  Finally, the y axis is perpendicular to the Galactic plane and completes a right-handed coordinate system.  

\begin{figure}[h!]
\begin{center}
\includegraphics[scale=0.85,angle=0]{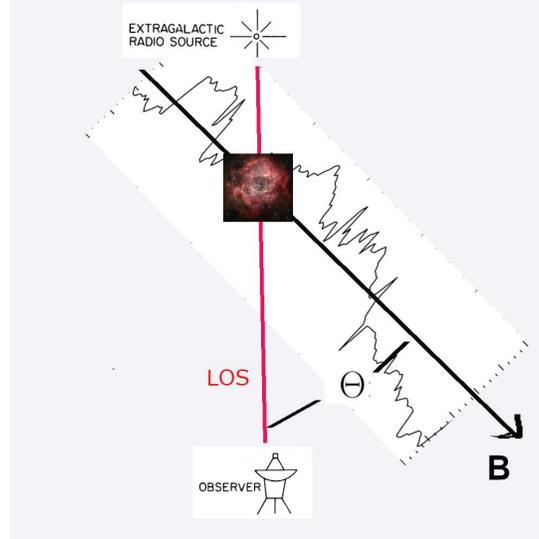}
\caption{Cartoon illustrating geometry of observer, polarized extragalactic radio source, and HII region.  A Galactic magnetic field is illustrated, consisting of a large scale, ordered component $\vec{B}_0$ inclined at an angle $\Theta$ with respect to the line of sight, and a superposed turbulent component $\vec{\delta b}$.  The plane of the paper is the Galactic plane, the z axis is defined by a ray from the radio source to the observer.  The coordinates x and y complete a right-handed coordinate system. In this illustration, only one component of the turbulent field is shown, $\delta b_p$, which is the component perpendicular to the large scale field and in the Galactic plane.  To lend realism, the trace shown represents 6 hours of turbulent solar wind magnetic fluctuations measured by the WIND spacecraft on September 25, 2020.  }
\end{center}
\end{figure}

The observational results of \citet{HarveySmith11, Savage13, Costa16, Costa18} show that there is a well-defined mean field throughout the HII regions, with a magnitude equal to or greater than the general Galactic field.  Given the known presence of plasma turbulence in the interstellar medium (ISM), we interpret this average HII region field to be the vector sum of the large scale galactic field $\vec{B_0}$ and a turbulent field $\vec{\delta b}$.  As indicated in Figure 2, the angle between the line-of-sight (LOS) and the large scale field is $\Theta$.  We also assume, for this analysis, that the large-scale field is in the Galactic plane.  

Given the configuration illustrated in Figure 2, we define a second coordinate system which is convenient for describing the turbulence.  The $\parallel$ direction is in the direction of the mean field, the $p$ direction is perpendicular to the mean field, and in the Galactic plane, and the $\perp$ direction completes the right-hand coordinate system and is perpendicular to the Galactic plane.  These three directions are defined by the unit vectors $\hat{e}_{\parallel}$,  $\hat{e}_p$, and $\hat{e}_{\perp}$. 

\subsection{Model for the Magnetic Field in the Vicinity of an HII Region}
 Given this latter coordinate system, we express the magnetic field (mean field plus turbulence) by 
\begin{equation}
\vec{B} = (B_0 + \delta b_{\parallel}) \hat{e}_{\parallel}  + \delta b_p \hat{e}_p  +  \delta b_{\perp} \hat{e}_{\perp}
\end{equation}
In analogy with the Alfv\`{e}nic turbulence in the solar wind, we expect $<(\delta b_p)^2 > = < (\delta b_{\perp})^2>$ and $<(\delta b_p)^2 > \gg < (\delta b_{\parallel})^2>$.  

The model we adopt is that $\delta b_{\parallel}, \delta b_p, \mbox{ and } \delta b_{\perp}$ are random, 0-mean quantities with an amplitude dominated by fluctuations on the outer scale l, $l \geq 2R$, where $R$ is the radius of the HII region.  In the cases of the Rosette Nebula \citep{Costa16} and W4 \citep{Costa18}, the values of $2R$ are 38 and 50 parsecs, respectively.  These values are comparable to, or slightly less than the outer scale of the 2D component of interstellar turbulence claimed by \citet{Minter96}.  

We adopt a very simple model for the average observed RM to a set of extragalactic radio sources viewed through an HII region, 
\begin{equation}
RM_{obs} = RM_{bck} + RM_{neb}
\end{equation}
where $RM_{bck}$ is the background RM due to the plasma in the Galaxy which would be observed in the absence of the nebula, and $RM_{neb}$ is the average RM for lines of sight through the HII region.  $RM_{neb}$ is given by the simple expression
\begin{equation}
RM_{neb} = 2 C n R A B_z
\end{equation}
where $C$ is the set of fundamental constants in curved brackets in Equation (1), $n$ is the mean electron density in the ionized portion of the HII region (determined by independent measurements such as intensity of thermal radio emission in \citet{Savage13, Costa16, Costa18}), $R$ again is the radius of the HII region, and $B_z$ is the z component of the interstellar field (mean field plus turbulence) at the location of the HII region.  
The variable $A \geq 1$ is a factor describing whether the magnetic field within the HII region is amplified over the value in the surrounding ISM.  \citet{HarveySmith11} assumed that it was not ($A = 1$), while \citet{Costa16, Costa18} considered physically-based models in which $A \leq 4$.  The case $A = 1$ was considered as well. 

$B_z$ is obtained from Equation (2) as 
\begin{equation}
B_z = \vec{B} \cdot \hat{e}_z
\end{equation}
Substituting Equation (2) into this expression and evaluating the dot products of unit vectors gives
\begin{equation}
B_z = (B_0 + \delta b_{\parallel}) \cos \Theta - \delta b_p \sin \Theta
\end{equation}
The $\delta b_{\perp}$ component of the turbulent field makes no contribution because, for lines of sight in or close to the Galactic plane, it is perpendicular to the LOS.  The negative sign in the second term arises because positive $\delta b_p$ is pointing away from the observer (Figure 2).  

Equation (6) shows the important, though obvious fact that the polarity of $B_z$ (i.e. the same as $\vec{B_0}$ or opposite) depends on both the amplitude of the fluctuations relative to the mean field as well as the angle $\Theta$ between the large scale field and the LOS.  

This expression can be rewritten in terms of dimensionless turbulence amplitudes (turbulent ``modulation indices'') as 
\begin{eqnarray}
B_z = B_0 ( \cos \Theta + x \cos \Theta - y \sin \Theta) \\
x \equiv \frac{\delta b_{\parallel}}{B_0} \\
 y \equiv \frac{\delta b_p} {B_0}
\end{eqnarray}

where $x,y$ are dimensionless, 0 mean random variables with assumed properties $\sqrt{<x^2>}, \sqrt{<y^2>} \leq 1$.  With all of this, the observed RM is then 

\begin{eqnarray}
RM_{obs} = RM_{bck} + 2CnRAB_0 (\cos \Theta + x \cos \Theta - y \sin \Theta) \\
RM_{obs} = RM_{bck} + RM_n (\cos \Theta + x \cos \Theta - y \sin \Theta)
\end{eqnarray}
with an obvious definition of the net nebular $RM_n$.  It is worth emphasizing that $RM_n$ is a property of the nebula itself, independent of the ISM or the geometry of the line of sight.  

The question considered in this paper then is whether the second term has the same sign as the first, and if not, if it is large enough to cause the observed RM to have the opposite sign as  $RM_{bck}$.  Equation (11) shows that this depends on two factors.  
\begin{enumerate}
\item The first is the relative magnitudes of $RM_{bck}$ and $RM_n$.  \citet{Savage13,Costa16,Costa18} found that for both the Rosette Nebula and W4, the largest absolute magnitudes of the RMs through the nebula were of the order of 1000-1500 rad/m$^2$, whereas the background Galactic RMs in those parts of the sky had absolute magnitudes of the order of 100 - 150 rad/m$^2$.  This indicates that the magnitude of $RM_n$ could be as high as 10 times that of $RM_{bck}$.
\item The statistics of the quantity
\begin{equation}
\xi = \cos \Theta + x \cos \Theta - y \sin \Theta
\end{equation}
are obviously crucial.  For the observed RM to be of the opposite polarity to $RM_{bck}$, it is necessary that $\xi < 0$, and in fact $\xi <  - \Delta$ where $\Delta$ is defined as the ratio of background to nebular RM.  
\end{enumerate}

For the remainder of this section, we turn to a discussion of the statistics of $\xi$.  

\subsection{Statistics of the Random Variable $\xi$}
We begin by making a further simplification and ignoring the x term in $\xi$.  The justification for this is the assumption stated above that for Alfv\`{e}nic turbulence, the fluctuations in the parallel component (to the mean field) are small compared to the transverse components.  In addition, both the mean field and the parallel component contributions to the stochastic variable $\xi$ are proportional to $\cos \Theta$.  We therefore proceed to consider the statistics of the variable 
\begin{equation}
\xi = \cos \Theta  - y \sin \Theta
\end{equation}

Only $y$ is a random variable, so we model it as a Gaussian-distributed, random variable with zero mean, with a probability distribution function
\begin{equation}
p(y) = \frac{1}{\sqrt{2 \pi} \sigma} \exp (-y^2 / 2 \sigma^2)
\end{equation}
The quantity $\sigma$ is the dimensionless amplitude of the fluctuations of the p component of the turbulence, and is a crucial parameter in our analysis.  

Given this expression we can generate the probability distribution function (pdf) of the stochastic variable $\xi$
\begin{equation}
p(\xi) = \frac{1}{\sqrt{2 \pi} \sigma} \frac{1}{\sin \Theta} \exp \left(-\frac{(\cos \Theta - \xi)^2 }{ 2 \sigma^2 \sin^2 \Theta} \right)
\end{equation}

For economy of notation, let $a \equiv \cos \Theta$, $b \equiv \sin \Theta$, with the obvious identity $b = \sqrt{1 - a^2}$, so that 
\begin{eqnarray}
p(\xi) = \frac{1}{\sqrt{2 \pi} \sigma b}  \exp \left( -\frac{(\xi - a)^2 }{ 2 \sigma^2 b^2} \right) \\
p(\xi) = \frac{1}{\sqrt{2 \pi} d}  \exp \left( -\frac{(\xi - a)^2 }{ 2 d^2} \right) 
\end{eqnarray}
with $d \equiv \sigma b$.  The distribution function of $\xi$ is also Gaussian, with a nonzero mean value  $< \xi> = a = \cos \Theta$.  

A sample plot of $p(\xi)$ is shown in Figure 2.  This figure shows that for sufficiently large-amplitude turbulence ($\sigma = 0.70$ in this case), and sufficiently large angles of the mean field with respect to the line of sight ($\Theta = 70^\circ$) the probability of the local field being reversed with respect to the mean field is substantial.  

\begin{figure}[h!]
\begin{center}
\includegraphics[scale=0.85,angle=0]{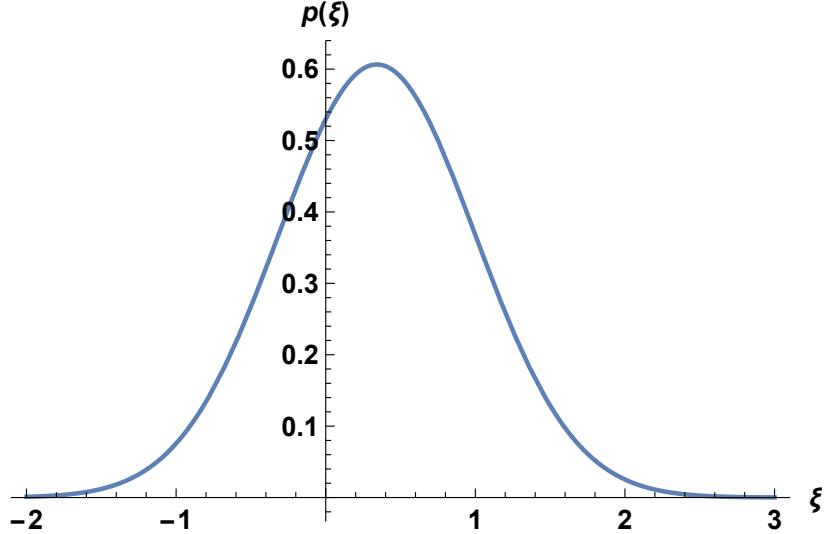}
\caption{The probability distribution function of the variable $\xi$  for the case in which the dimensionless amplitude $\sigma$ of the magnetic fluctuations in the p component of the magnetic field is $\sigma = 0.70$ and the angle $\Theta$ between the line of sight and the mean magnetic field is 70 degrees.  In this case, the probability of $\xi$ being negative is substantial.}
\end{center}
\end{figure}

\subsection{Expression for $P_-$, the Probability of a Polarity Reversal}
Given the expression in Equation (16) or (17) for the probability distribution function of the parameter $\xi$, we have the quantity of interest, the total probability that the line-of-sight component of the magnetic field at a given location in the ISM will be reversed from that of the mean field, We note this quantity by $P_-$, and the obvious expression for it is 
\begin{equation}
P_- = \int_{-\infty}^{- \Delta} d \xi p(\xi)
\end{equation} 

By inspection, it can be seen that this integral is the same as 

\begin{equation}
P_- = \int_{-\infty}^{- (\Delta +a)} d y p_G(y,d) =  \int_{\Delta +a}^{\infty} d y p_G(y,d)
\end{equation} 
where $p_G(y,d)$ is a zero mean, normalized Gaussian pdf with standard deviation $d$.  

If we make a final change of variables, $y \rightarrow t \equiv \frac{y}{\sqrt{2} d}$, then Equation (19) becomes 
\begin{eqnarray}
P_- = \frac{1}{\sqrt{\pi}} \int_{X}^{\infty} dt e^{-t^2} \\
X \equiv \frac{\Delta + a}{\sqrt{2} d}  = \frac{(\Delta + \cos \Theta)}{\sqrt{2} \sigma \sin \Theta}
\end{eqnarray} 
Equation (20) is very close to the standard form of the Error Function \citep{Beckmann67} 
\begin{equation}
erf(X) \equiv \frac{2}{\sqrt{\pi}}  \int_0^{X} dt e^{-t^2}
\end{equation}

Comparison of Equations (20) and (22) allows the final, compact formula for the total probability of polarity reversal, 
\begin{equation}
P_- = \frac{1}{2} \left[ 1 - erf(X) \right] = \frac{1}{2} erfc(X)
\end{equation}
where $erfc(X)$ is the complement of the error function.  

Equation (23) is plotted in Figure 4.  
\begin{figure}[h!]
\begin{center}
\includegraphics[scale=0.85,angle=0]{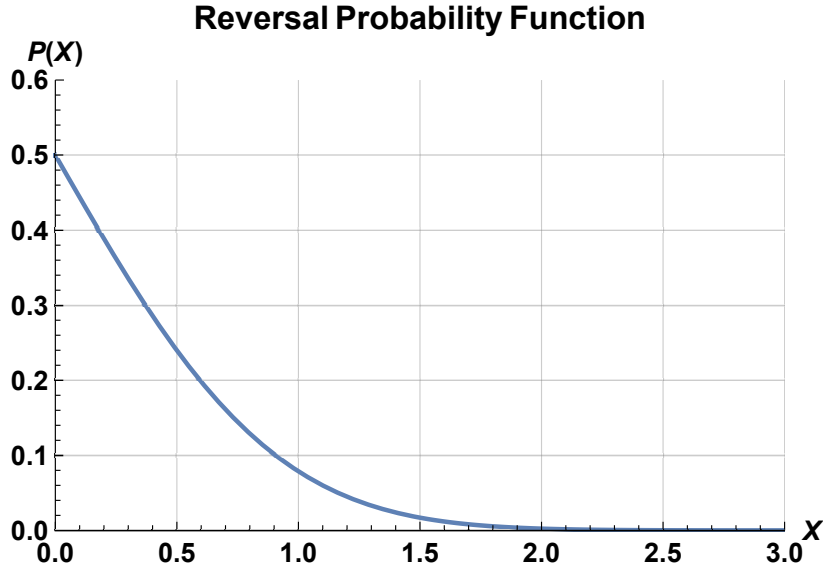}
\caption{The probability that the line of sight component of the magnetic field at the location of an HII region is opposite to that of the global, Galactic field.  The probability $P_{-}$ is a function of the parameter $ X = \frac{(\Delta + \cos \Theta)}{\sqrt{2} \sigma \sin \Theta}$ }.
\end{center}
\end{figure}
As the argument $X \rightarrow 0$, the probability of reversal $P_- \rightarrow 0.5$.  This makes complete sense, since small argument $X$ corresponds to strong turbulence and/or a mean field nearly perpendicular to the line of sight.  In this case, the local magnetic field is dominated by the turbulent component $\delta b_p$, which has a 50 percent change of being positive and and a 50 percent chance of being negative.  

The case of $X \rightarrow \infty$ corresponds to weak turbulence, and/or a mean field increasingly directed towards the observer.  In this case, the probability of a reversal in the line-of-sight component of the magnetic field approaches zero.  

The probability of a reversal in the line of sight component is determined by the parameter $X$, and it is of interest to calculate this quantity for some observationally-relevant cases.  

\subsection{Calculations Relevant to Observations of HII Regions}
In this section, we briefly apply these formulas to the observations presented in Section 2.  A total of 6 HII regions distributed over a range of Galactic longitudes precludes a convincing and sophisticated statistical analysis.  However, the results presented in Figure 1 suggest that theoretical values of $P_- \geq 0.20$ would begin to have an uncomfortable confrontation with the observations.  On the other hand, if characteristics of the HII regions, amplitude of ISM turbulence and direction of the line of sight indicate a value of $P_- \sim 0.01$, our calculations are completely consistent with the observations discussed in Section 2.  The formula given in Eq. (21) shows that the parameter $X$ depends on $\Theta$, $\sigma$, and $\Delta$.  In what follows, we adopt a value of $\Delta = 0.1$, which is in accord with the observations of the Rosette Nebula and W4 discussed above.  

\citet{Savage13,Costa16}  in their study of the Rosette nebula, found that reasonable fits of simple magnetized shell models to the set of RM measurements gave a value of $\Theta \simeq 70^{\circ}$ if the interstellar magnetic field is amplified in the shell by shock compression, and $\Theta \simeq 50^{\circ}$ if there were no change in the magnitude of the magnetic field in the shell, with respect to the general ISM \citep[see Table 5 of ][]{Costa16}.  \citet{Savage13} noted that an independent estimate, based on models of the Galactic magnetic field, would give $\Theta \simeq 60^{\circ}$ at the location of the Rosette.  \citet{Costa18} used a value of $\Theta = 55^{\circ}$ for W4, if the Galactic magnetic field is approximated as azimuthal.  Thus values of $50^{\circ} \leq \Theta \leq 70^{\circ}$ are relevant to the results shown in Figure 1.  

The amplitude of the turbulent magnetic field on spatial scales of tens of parsecs is not well known, but results from the literature discussed in Section 1 indicate that it is of order the mean, large scale field. Thus our parameter $\sigma$ if probably of the order of unity, or a large fraction thereof.  A value of $\sigma =  0.70 \simeq \frac{1}{\sqrt 2}$ would have the magnitude of fluctuating field (quadratic sum of $ \delta b_p$ and $\delta b_{\perp}$ components) equal to the magnitude of the large scale field.  

A value of $\sigma = 0.70$ and $\Theta = 70^{\circ}$ gives $X = 0.48$; reference to Figure 4 shows that the probability of reversal $P_-$ is about 24 \%.  In such a case it is mildly surprising that all 6 HII regions discussed in Section 2 produce ``Faraday rotation anomalies'' with the same polarity as the large scale field.  On the other hand, a slightly smaller value of $\sigma = 0.50$ and  $\Theta = 50^{\circ}$, also consistent with the \citet{Costa16} results for the Rosette Nebula and W4, would have  $X = 1.37$.  In this case, Figure 4 shows that the probability of a polarity reversal for a single source is less than 5 \%, and the results of Figure 1 are unremarkable.  

This discussion shows that future investigations on more HII regions could provide more stringent limits on the large scale turbulent fluctuations in the Galactic magnetic field.  The formulas presented in this section provide a mathematical vocabulary for discussing current as well as future observations.  For example, it is clear that the probability of a reversal becomes much greater for $\Theta \sim \frac{\pi}{2}$.   HII regions in the close vicinity of the  Galactic anticenter would provide particularly sensitive diagnostics for the amplitude of large scale turbulence.  

\section{4. Conclusions from Zeeman Effect Measurements of Star Formation Region Masers}
The analysis in Section 3 assumes that HII regions occur in the presence of an unmodified Galactic magnetic field.  In the simplified stellar bubble models presented in \citet{Savage13, Costa16, Costa18} the expanding bubble modifies the Galactic field, but the RM enhancement due to the bubble is proportional to the Galactic field (mean field plus turbulent contribution at the location of the HII region) in the absence of the HII region.  In the model considered by \citet{HarveySmith11} the HII region consists of an ionization of the interstellar gas with no modification of the Galactic field.  \citet{HarveySmith11} describe this as the HII region ``lighting up'' the Galactic field that exists at the location of the HII region.  

These viewpoints may be physically naive.  HII regions are produced by hot, massive stars that are products of the star formation process. Star formation regions, in turn, result from contraction of gas from the general ISM, its conversion to molecular form, and an associated increase in gas density.  It therefore seems likely that the process of forming dense clumps in molecular clouds from which stars form results in major modification of the Galactic field, perhaps to the extent of totally randomizing the vector field from its value in the absence of the star formation region.  In this light, the observational results presented in Section 2 are even more remarkable; the preservation of the polarity of the large scale field occurs in spite of both the presence of natural turbulence in the Galactic field, as well as the possibly major variations to the field induced by the star formation process.  

The HII region results presented in Section 2 are not the only ``local''  measurements of the magnetic field in the ISM.  Hydroxyl, methanol, and water masers in star formation regions display the Zeeman effect, permitting measurements of the magnitude and sign of the LOS component of the magnetic field in the maser region \citep{Crutcher19}.  These masers occur in regions of very high density compared to the ISM far from star formation regions, and thus probe gas that has been substantially modified in density, and presumably magnetic field, by the star formation process \citep{Crutcher19}.  

In spite of this, observations of the polarity of the magnetic field in maser regions show results similar to those presented in Section 2.  \citet{Fish03} examined Zeeman effect measurements for more than 50 massive star formation regions.  Although they did not find a general Galactic-wide magnetic field direction, they did find that `` \ldots in the solar neighborhood the magnetic field outside the solar circle is oriented clockwise as viewed from the north Galactic pole \ldots''.  This is the same sense as shown in Figure 1 on the basis of the HII region results.  A subsequent study by \citet{Green12} yielded similar results.  \citet{Green12} reported measurements of the magnetic field (sign and magnitude of the line-of-sight component of the magnetic field) for 14 Zeeman pairs in 6 high mass star formation regions in the Carina-Sagittarius arm (4th Galactic quadrant).  Their results show a dominant magnetic polarity in this spiral arm (See Figure 7 of \citet{Green12} .  The significance of these observational results was summarized in \citet{Han17}, who noted that they implied  ``turbulence in molecular clouds and violent star-forming regions seem not to alter the mean magnetic field direction, although the field strength is much enhanced from a few $\mu$ G to a few mG ''. 

\section{5. Speculations from the Study of Hydrodynamical Turbulence}
The results for the HII regions presented above, and illustrated in Figure 1 indicate, at the very least, potentially significant limits on the magnitude of turbulent magnetic field fluctuations in the Galactic field.  The results on star formation region OH masers from \citet{Fish03} and \citet{Green12} indicate, and the HII regions results may indicate, a far stronger and mysterious physical process: a memory of the large scale polarity of the Galactic field as it is modified during the contraction and increase in density associated with star formation.  

The obvious question which then arises is the meaning of  ``the persistence of memory'' in the large scale, Galactic magnetic field.  The problem arises because of the assumed isotropic nature of turbulent fluctuations on scales much smaller than the outer scale of the turbulence.  Such fluctuations, deep in the inertial subrange or close to the dissipation range, should have no tendency to respect large scale asymmetries in the Galactic plasma. It is worthwhile to look for insight from the study of hydrodynamical turbulence.  There are obvious physical differences between  the turbulence in the diffuse plasmas of the interstellar medium and the turbulence in gases and liquids in laboratory and industrial settings.  However, there are important similarities, too, and both the quality of the diagnostics and the shear amount of effort expended in the study of hydrodynamic turbulence exceeds that which is possible for astrophysical turbulence.  

A review of our understanding of ``passive scalars'' in hydrodynamic turbulence is presented in \citet{Warhaft00}. Passive scalars are quantities which are convected by fluid turbulence, without affecting the flow field.  \citet{Warhaft00} considered as a prime example temperature fluctuations in hydrodynamic turbulence. An astrophysical example of a passive scalar could be the plasma density fluctuations responsible for interstellar radio scintillations.  \citet{Warhaft00} presents an illuminating discussion of the differences between the engineering and physics views of turbulence, and offers an admonition that is certainly relevant for Galactic astronomers as well: `` The physicists, with their quest for universality, have concentrated on inertial and dissipation scales, the hope being, if the Reynolds (Re) and Peclet (Pe) numbers are high enough, that the small-scale behavior will be independent of the large scales.  We address both the large and small scales in this review and argue that this division is artificial and dangerous''.  \citet{Warhaft00} then goes to present laboratory and theoretical evidence that small scale passive scalar fluctuations (primarily temperature fluctuations) are anisotropic with an anisotropy that reflects large scale asymmetry.  In the cases discussed by \citet{Warhaft00} this large scale asymmetry was most typically a transverse temperature gradient imposed on a turbulent shear flow.  \citet{Warhaft00} also shows that the small scale passive scalar fluctuations possess a skewness and kurtosis that exceeds that present in the fundamental turbulent velocity field.  

It is not clear to me if the clear and insightful presentation of \citet{Warhaft00} holds a key to understanding the mysterious observation described in this paper, or if any such comparison would be a hallucinogenic extrapolation of laboratory results far outside their range of applicability. While Warhaft's review does describe a documented persistence of large scale anisotropy on small turbulent scales, as appears to be indicated in our Galactic ISM observations, it must be emphasized that the discussion of \citet{Warhaft00} is concerned with passive scalars in the turbulence. In the Galactic ISM context, that would be plasma density, temperature, ionization fraction, and metallicity.  The magnetic field, on the other hand, should share status with fluid velocity as a ``primary excitation'' of the turbulence, if the ISM turbulence is Alfv\'{e}nic.  Nonetheless, it would be worthwhile to consider further if the extensive, laboratory based studies of turbulence from an engineering perspective can illuminate the nature of plasma turbulence in the interstellar medium.

\section{6. Summary and Conclusions}   
\begin{enumerate}
\item In the case of the Galactic-anticenter-direction HII regions Rosette Nebula and W4, the mean magnetic field in the HII region has the same polarity as the general Galactic field in that part of space, despite the presence of strong MHD turbulence which would be expected to randomize the polarity.  An additional 4 HII regions studied by Harvey-Smith et al (2011) share this property.  
\item We derive formulas for the probability of the ``point'' magnetic field having the same polarity as the general field, which depends on properties of the turbulent magnetic fluctuations as well as the details of the line of sight.  For reasonable estimates, this probability could be non-negligible, being as high as 25 \% for plausible characteristics.  As a result, the observation of 6 anticenter-direction HII regions, all having the same magnetic polarity of the general field, becomes mildly curious.  
\item The results on HII regions support similar results on the polarity of star formation OH masers presented by \citet{Fish03} and \citet{Green12}.  In those studies, the polarity of the magnetic field in maser regions is correlated among star formation regions that are spatially separated, and the polarity is apparently determined by that of the large scale Galactic field.  The maser region results not only limit the properties of turbulent fluctuations in the Galactic field, but also restrict the field randomization that presumably occurs in the assembly of star formation cores during the process of star formation.  
\item We point out that studies of turbulence in laboratory settings also show a persistence of large scale asymmetries in small and inertial scale fluctuations, which could be termed ``the persistence of memory'' in the Galactic magnetic field.  These results are discussed in \citet{Warhaft00}, and refer to passive scalars in hydrodynamic turbulence.  Although it is unclear if the results discussed in \citet{Warhaft00} are applicable to Galactic MHD turbulence, it would appear to be a worthwhile path to investigate.  

\end{enumerate}

\appendix
\section{Glossary of Mathematical Variables}
\begin{deluxetable}{|c|l|}[h]
 \tablehead{\colhead{Variable} & \colhead{Definition}} 
\startdata
$\Delta \chi$ &  change in polarization position angle due to Faraday rotation (Eq.1)\\
$\vec{B}$ & total magnetic field in ISM at location of HII region (Eq.2)\\
$\vec{B}_0$ & large scale, ordered component of Galactic magnetic field\\
$\vec{\delta b}$ & turbulent component of Galactic magnetic field\\
$\Theta$ & angle between LOS and large scale magnetic field $\vec{B_0}$ at location of HII region (Fig. 2) \\
$(\hat{e}_{\parallel}, \hat{e}_p, \hat{e}_{\perp})$ & unit vectors defining coordinate system aligned with large scale field $\vec{B_0}$\\
$(\delta b_{\parallel}, \delta b_p, \delta b_{\perp})$ & components of the turbulent magnetic field in a coordinate system aligned with large scale field $\vec{B_0}$\\
$RM_{obs}$ & observed RM for a background source observed through an HII region (Eq.3)\\
$RM_{bck}$ & RM due to ISM along LOS, but excluding HII region (Eq.3)\\
$RM_{neb}$ & RM due to plasma in HII region (Eq.3)\\
$C$ & set of atomic constants appearing in expression for Faraday rotation (Eq. 1)\\
$n$ & electron density in plasma causing Faraday rotation (Eq. 1)\\
$R$ & radius of HII region\\
$B_z$ & LOS component of Galactic magnetic field at location of HII region\\
$A$ & factor ($\geq 1$) determining whether Galactic field is amplified within HII region.  Model dependent (Eq. 4)\\
$\hat{e}_z$ & unit vector in direction of LOS\\
$x$ & dimensionless amplitude of turbulent magnetic fluctuations in the direction of the large scale field (Eq. 8)\\
$y$ & dimensionless amplitude of turbulent magnetic fluctuations $\perp$ to large scale field (Eq. 9)\\
$\xi$ & derived variable dependent on $\Theta$, $x$, and $y$ (Eq. 12)\\
$\Delta$ & threshold value of $\xi$ such that sign of RM reversed for $\xi < -\Delta$\\
$\sigma$ & rms value of Gaussian-distributed $y$ (Eq. 14)\\
$a$ & secondary variable $a \equiv \cos \Theta$ (Eq. 16)\\
$b$ & secondary variable $b \equiv \sin \Theta$ (Eq. 16)\\
$d$ & secondary variable $d \equiv \sigma b$ (Eq. 17)\\
$p(\xi)$ & normalized, Gaussian pdf of $\xi$ (Eq. 15-17)\\
$p_G(\xi)$ & shifted, zero-mean version of $p(\xi)$ (Eq. 19)\\
$P_-$ & probability that turbulent fluctuations cause reversal of LOS magnetic polarity (Eq. 18)\\
$t$ & secondary variable $t \equiv \frac{y}{\sqrt{2} d}$ (Eq. 17)\\
$X$ & single parameter determining $P_-$ (Eq. 21)\\
\enddata
\end{deluxetable}

\end{document}